\documentstyle[preprint,aps]{revtex}
\begin{document}

\begin{title} 
{        A Critique of ``A Critique of Two Metals''
}
\end{title}

\author{ Philip W. Anderson and G. Baskaran}

\address {Joseph Henry Laboratories of Physics\\
Princeton University, Princeton, NJ 08544}
\maketitle
\vskip.5truein
\baselineskip=18pt

  The ``Critique''\cite{1} contains in its first few paragraphs an elegant,
if somewhat incorrect, statement of the issues between us and the
school which believes, almost religiously, in the quantum
critical point as the solution to all our woes in the cuprates.

  The fundamental argument is presented in the second paragraph:
``Ten years of work by some of the best minds in theoretical
physics have failed to produce any formal demonstration''\dots of the
Mott insulating state. The statement would be ludicrous if it
were not so influential. The proviso ``at zero temperature'' is
added, because of course most Mott insulators order magnetically
at some finite, if often low, temperature; the Mott insulator is
not a zero-temperature fixed point, in general. Neither, for that
matter, is the Fermi liquid.  But one does not need a formal
demonstration--although I believe I provided that, if after
Mott's original papers that was necessary, in my 1959 paper. The
world, if one lifts one's eyes from the computer screen, is full
of examples, and I believe that one concrete, material example is
worth a million hours of computer time. Two which are very
relevant to the case in point are $CuSO_4\cdot 5H_2O$, or blue vitriol to
our ancestors; and $CuCl_2\cdot 2H_2O$. Both are examples of
$Cu^{(++)}$ and
are not only insulating but transparent with a beautiful blue
color, at all reasonable temperatures---they deliquesce if you get
them too hot. The chloride was an elegant demonstration case for
antiferromagnetic spin waves below its $He$-temperature Ne\'el point;
the sulphate was an early subject of adiabatic demag studies by
Laughlin's colleague T. Geballe, and as far as I know is
paramagnetic down to very low temperatures. Some other 
 less perfect cases are very important to us---hemoglobin,
which in its liquid form is familiar to all of us; and the three
or four oxides of iron---rust, which is mostly goethite; hematite,
of which there are happily mountains; and magnetite, known to the
ancients on both continents and just to show that the ground
state doesn't always turn out antiferromagnetic.

  As I think Laughlin must know, the Mott insulator is a form of
quantum solid, and the melting transition in He3 is our best
example of a Mott transition. Our objection to trying to fit
cuprates into a quantum critical point scenario in the way that
Zhang does becomes obvious when one tries to do the same with
$p$-wave superconductivity and antiferromagnetic $^3He$ solid. There
are similarities which can be exploited between the short-range
correlations of the quantum solid and the quantum liquid, but no
connection in terms of symmetry and asymptotics. It is well-known
that no critical point, even in classical theory, connects solid
and liquid. The Mott transition (as is seen in $V_2O_3$) is a
first-order line ending in a critical point, classically, but
this implies nothing about any relationship between the two
phases at low temperatures.

  I am sorry to belabor the point that there is a well-defined
insulating state in which the degrees of freedom are spins only,
with an energy gap to charged excitations; and trying to connect
this high-temperature (relatively) state continuously to a metallic
state by some smooth transformation does not make any physical
sense. But this seems to be unfamiliar to the generation of
physicists who did not grow up with paramagnetic resonance as a
major concern. It is the tragedy of Mott that although he almost
certainly won his Nobel prize for the Mott insulator, Slater, who
couldn't think clearly about finite temperature, won the
publicity battle. 

  In the first paragraph of the Critique the content and intent of Baskaran
and my discussion is confused with our opinions on the source of
superconductivity in the cuprates. Our objection to the kind of
quantum critical point suggested by Zhang has nothing to do with
whether it connects to a Fermi liquid or a non-Fermi liquid; our
statement is that whatever the metallic state is, the low-energy
excitations must
be described in terms of a Fermi surface, that is a surface in
momentum space which is the locus of all of the single-particle
amplitude, and which encloses a finite volume. Unlike the
relativistic field theories and critical point theories with
which Zhang is familiar, the excitation spectrum does not derive
from fluctuations of a field which is uniform in space. The order
parameter which characterizes the generalized fermi liquid state
is this surface, and its fluctuations are the bosonic excitations
from which quasiparticles can be constructed as solitons.
The theory when bosonized thus has the 
kind of structure described by Haldane and
others, involving fluctuations of a surface in momentum space.
This description is actually equally valid whether the resulting
theory is FL or NFL. It has been the most serious difficulty of
the school which has attempted to bring the cuprates under the
aegis of one form or another of field theory---usually gauge
theory---that the forms of the theories they used were not yet
sufficiently advanced to deal with the Fermi surface, which is
obvious in all the experimental manifestations of these
materials.

  These two underlying intermediate-energy states are
incompatible in every way. The superconductor derives from a
Fermi surface---experimentally. As Campuzano, Norman and
co-workers show, the minimum gap is always at the Luttinger Fermi
surface. The antiferromagnet derives from a Mott insulator. The
two are immiscible and many complex phenomena---such as
Stripes---are found in the unstable two-phase region between them.
(Mott in 1956 described this fundamental instability in terms of
the impossibility of adding a small number of free carriers in
the magnetic case.)

Finally, we object to the statement that Zhang articulates ``an
alternate view in a particularly simple and elegant way
$.\,.\,.\,.\,.$ that
everyone can understand''.  We, for one, find that the
presentation, while extremely smooth, is not in any way
understandable, since it is expressed in terms which we cannot
accept as having relevance to the problem, using buzz-words which
relate to elegant---but not particularly fruitful---treatments of
critical points without reference to the actual physical content.
The ``anyone'' certainly does not refer to us nor to any
experimentalist in the field with whom we are familiar; and I
hope that there are theorists also who can see through a
non-existent set of clothes.  

  The remainder of the ``Critique'' is not directed primarily at
our discussion of Zhang's paper at all,  but aside from some
rather immoderately phrased criticisms of our work based on the
rather irrelevant point that the Mott state does not exist in
some very restricted sense which Laughlin chooses to define, it
seems to be presenting a 
new or revised version of Laughlin's own theory, so does not
require our answer. Laughlin seems to be declaring closed a
series of discussions of which I am sure few of the discussants
would consider themselves ready to terminate in these terms. 

  One point is worth making in the context of a discussion of
critical lines and crossovers. There is a crossover line
associated with the High $T_c$ phenomenon, which might be thought of
as concealing an underlying zero-temperature critical point.
This is the crossover between two- and three-dimensional
metallicity. One can hardly doubt that the great majority of High
$T_c$'s show only incoherent transport along the $c$-direction in the
normal state. It is
also clear that they are all three-dimensional superconductors
with $c$-axis supercurrents, hence coherence in the c-direction. As
with all metal-insulator transitions(see above) it is not
possible to define the insulator unequivocally except at $T=0$,
hence there is the presumption of a quantum critical point. But
superconductivity intervenes. The $c$-axis infrared data
demonstrate these phenomena so beautifully that it is hard to see
how so many theorists can ignore the role of the third dimension.
Once one is overdoped, the two-dimensionality is gone as is $T_c$.

\end{document}